\documentclass[twoside]{mhd}

\mhdhead{40}{1}{1}	

\usepackage{graphicx}
\newcommand{\be}{\begin{equation}}
\newcommand{\ee}{\end{equation}}
\newcommand{\bea}{\begin{eqnarray}}
\newcommand{\eea}{\end{eqnarray}}
\newcommand{\ba}{\begin{array}}
\newcommand{\ea}{\end{array}}
\newcommand{\bi}{\begin{itemize}}
\newcommand{\ei}{\end{itemize}}
\newcommand{\ben}{\begin{enumerate}}
\newcommand{\een}{\end{enumerate}}
\newcommand{\etal}{{\em et al.}}

\title{Energy transfers in MHD turbulence and its applications to dynamo}

\author{Mahendra K. Verma\inst{1}, Rodion Stapanov\inst{2},
Franck Plunian\inst{3}}

\institute{ Department of Physics, Indian Institute of Technology Kanpur, Kanpur 20816, India
\and
Institute of Continuous Media Mechanics, Korolyov 1, 614013 Perm, Russia
\and
Laboratoire de G\'{e}ophysique Interne et Tectonophysique, Universit\'{e} Joseph Fourier, CNRS, Maison des G\'{e}osciences, B.P. 53, 38041 Grenoble Cedex 9, France} 

\begin{document}
\maketitle


\begin{abstract}
In this paper we describe the mode-to-mode energy transfers and energy fluxes of MHD turbulence.  These energy transfers are very useful for understanding the turbulence dynamics, as well as for applications such as dynamo.  We illustrate how the energy fluxes provide valuable insights into the mechanism of growth of the large-scale magnetic energy in dynamo.
\end{abstract}


\section{Introduction}
Turbulence is highly intractable problem, primary due to the strong nonlinearity.   The most well-known result of hydrodynamic turbulence is by Kolmgorov~\cite{Kolmogorov:DANS1941Dissipation}.  According to this theory, when the fluid is forced at large scales, a constant energy flux $\Pi$ flows from large scales to small scales via intermediate scale.   In the intermediate scale, the energy spectrum is $E(k) = K_\mathrm{Ko} \Pi^{2/3} k^{-5/3}$, where $K_\mathrm{Ko}$ Kolmogorov's constant.

MHD turbulence is more complex than hydrodynamic turbulence due to more number of variables (velocity and magnetic fields) and parameters (viscosity and magnetic diffusivity).  Also, the equations of MHD turbulence has four nonlinear terms that induces more kinds of energy fluxes.  In this paper we describe these fluxes for MHD turbulence.  Also, these fluxes are best derived using ``mode-to-mode energy transfers"~\cite{Dar:PD2001, Verma:PR2004}, which will also be discussed in this paper.

The outline of the paper is as follows: In Sections~\ref{sec:governing_eqns} and \ref{sec:energy_eqns} the dynamical equations and energy equations respectively.  The mode-to-mode energy transfers as well as their physical interpretations are described Sec.~\ref{sec:energy_eqns}.  In Sec.~\ref{sec:energy_fluxes} we discuss the energy fluxes in MHD turbulence.  Sec.~\ref{sec:dynamo} contains an application of energy fluxes to dynamo.  We conclude in Sec.~\ref{sec:conclusions}.

\section{Governing equations:}
\label{sec:governing_eqns}

The equations of incompressible magnetohydrodynamics are
 \bea 
\frac{\partial{\mathbf{u}}}{\partial{t}}  + ({\bf u} \cdot \nabla) {\bf u} & = &  -  \nabla p  + ( {\bf B} \cdot \nabla) {\bf B} + \nu \nabla^2 {\bf u} + {\bf F}_\mathrm{ext}, \label{eq:MHD_formalism:MHDu} \\
\frac{\partial{\mathbf{B}}}{\partial{t}} + ({\bf u} \cdot \nabla) {\bf B}   & = &  ({\bf B} \cdot \nabla) {\bf u}   + \eta \nabla^2 {\bf B},  \label{eq:MHD_formalism:MHDB} \\
\nabla \cdot {\bf u} = \nabla \cdot {\bf B} & = & 0, 
\label{eq:MHD_formalism:MHDdel_uB_zero}  
\eea
where ${\bf u, B}, p$ are respectively the velocity, magnetic, and pressure fields, $\nu$ is the kinematic viscosity, and $\eta$ is the magnetic diffusivity.   Note that the pressure $p$ is sum of hydrodynamic and magnetic pressures, and ${\bf B}$ could include a constant magnetic field (corresponding to wavenumber ${\bf k}=0$).   $ {\bf F}_\mathrm{ext}$ is the external force field, which is typically employed at large scales.   

In Fourier space, the above equations  are transformed to
 \bea
  \frac{d}{d t}  {\bf u} (\mathbf{k}) + {\bf N}_u ({\bf k})
& = & -i {\bf k} p (\mathbf{k}) +{\bf F}_u({\bf k})  -  \nu k^{2}  {\bf u}(\mathbf{k}) + {\bf F}_\mathrm{ext}({\bf k}), \label{eq:MHD_formalism:MHDuk} \\
  \frac{d}{d t}  {\bf B} (\mathbf{k}) + {\bf N}_B ({\bf k})
& = &  {\bf F}_B({\bf k})  -  \eta k^{2}  {\bf B}(\mathbf{k}) , \label{eq:MHD_formalism:MHDBk}\\
{\bf k\cdot u}(\mathbf{k}) = {\bf k\cdot B}(\mathbf{k}) & = & 0, \label{eq:MHD_formalism:k_uBk_zero} 
\eea 
where the nonlinear terms are 
\bea 
{\bf N}_u ({\bf k})= i  \sum_{\bf p}  {\bf \{ k \cdot u(q) \} u(p)};~~~ 
{\bf N}_B ({\bf k}) =  i  \sum_{\bf p}  {\bf \{ k \cdot u(q) \} B(p)} ,\label{eq:MHD_formalism:MHD_NBk}  \\
{\bf F}_u ({\bf k})= i  \sum_{\bf p}  {\bf  \{ k \cdot B(q) \} B(p)};~~~{\bf F}_B ({\bf k})  = i  \sum_{\bf p}  {\bf  \{ k \cdot B(q) \} u(p)}, \label{eq:MHD_formalism:MHD_NBk}  
\eea
where ${\bf k=p+q}$. The terms ${\bf N}_u ({\bf k}) , {\bf N}_B ({\bf k}) $ represent the advection of the velocity and magnetic fields by the velocity fluctuation ${\bf u(q)}$, while ${\bf F}_u $ is the Lorentz force, and ${\bf F}_B$ corresponds to the  nonlinear term ${\bf (B \cdot \nabla) u}$.  The pressure is determined using
\be
 p (\mathbf{k}) = i \frac{1}{k^2}{\bf k} \cdot [{\bf N}_u(\mathbf{k}) - {\bf F}_u(\mathbf{k})].
 \ee
 We assume that ${\bf F}_\mathrm{ext}({\bf k})$ is divergdence-free.
 
 We also define the modal kinetic and magnetic energies using the following formulas:
 \be
E_u(\mathbf{k})  = \frac{1}{2} |{\bf u(k)}|^2;~~~E_b(\mathbf{k})  = \frac{1}{2} |{\bf b(k)}|^2.
\ee
 In the next section we describe the  equations for the kinetic and magnetic energies, and derive formulas for the mode-to-mode energy transfers   for MHD turbulence.

\section{Energy equations:}
\label{sec:energy_eqns}

In real space, the equations for the kinetic energy density, $E_u({\bf r})$, and magnetic energy density, $E_b({\bf r})$, are as follows:
\bea
\frac{\partial}{\partial t} E_u({\bf r})  + u_j  \partial_j  \left(\frac{1}{2} u_i u_i \right)   & =  & - \partial_j (u_j p) + [ B_j \partial_j  B_j] u_i  - \nu   \omega^2, \label{eq:MHD_formalism:Eu_dynamics} \\
\frac{\partial}{\partial t} E_b({\bf r})  + u_j  \partial_j  \left(\frac{1}{2} B_i B_i  \right)   & =  &    [B_j   \partial_j   u_i ] B_i - \eta   J^2, \label{eq:MHD_formalism:Eb_dynamics} 
\eea
where $\omega = \nabla \times {\bf u}$ is the vorticity, and ${\bf J = \nabla \times B}$ is the current density.  To derive the corresponding equations in Fourier space, we perform scalar product of Eq.~(\ref{eq:MHD_formalism:MHDuk}) with ${\bf u^*(k)}$; the resulting equation and its complex conjugate are added together to obtain an equation for the modal kinetic energy.  Similar operations on  Eq.~(\ref{eq:MHD_formalism:MHDBk}) with ${\bf B^*(k)}$ yields the equation for the modal magnetic energy. These equations are 
\bea
 \frac{d}{dt} E_u(\mathbf{k}) & = &  - \Re[ {\bf N}_u({\bf k}) \cdot {\bf u}^*({\bf k}) ] +    \Re[ {\bf F}_u({\bf k}) \cdot {\bf u}^*({\bf k}) ]   \nonumber \\ 
 & = &  \sum_{\bf p} \Im \left[ {\bf \{  k \cdot u(q) \}  \{ u (p) \cdot  u^*(k) \}  - \{  k \cdot B(q) \}  \{ B (p) \cdot  u^*(k) \}}  \right] ,
 \label{eq:MHD_formalism:Euk}  \nonumber \\  \\
  \frac{d}{dt} E_B(\mathbf{k}) & = &  - \Re[ {\bf N}_B({\bf k}) \cdot {\bf B}^*({\bf k}) ] +    \Re[ {\bf F}_B({\bf k}) \cdot {\bf B}^*({\bf k}) ]    \nonumber \\
 & = &   \sum_{\bf p} \Im \left[ {\bf \{  k \cdot u(q) \}  \{ B (p) \cdot  B^*(k) \}  - \{  k \cdot B(q) \}  \{ u (p) \cdot  B^*(k) \}}  \right]  ,  \nonumber \\
 \label{eq:MHD_formalism:EBk} 
  \eea
where ${\bf k=p+q}$.  For simplification we set $\nu=\eta=0$.  Note that the pressure $p({\bf k})$ disappears due to the incompressibility condition ${\bf k \cdot u(k)} = 0$.

\subsection{Combined energy transfers in MHD turbulence:}

The sums in Eq.~(\ref{eq:MHD_formalism:Euk}, \ref{eq:MHD_formalism:EBk}) involves all the modes of the system.  However, considerable insights are gained when we focus on  a pair of triads, $({\bf X,Y,Z})$ and  $({\bf -X,-Y,-Z})$.  Note that ${\bf X+Y+Z}=0$ for  a triad. Under this truncation, the above energy equations get reduced to the following form:
\bea
 \frac{d}{dt} E_u(\mathbf{X}) & = & S^{uu}({\bf X|Y,Z}) + S^{ub}({\bf X|Y,Z}),
  \label{eq:MHD_ET:Euk_triad} \\
   \frac{d}{dt} E_b(\mathbf{X}) & = & S^{bb}({\bf X|Y,Z}) + S^{bu}({\bf X|Y,Z}),
  \label{eq:MHD_ET:Ebk_triad}
 \eea
 where
 \bea
 S^{uu}({\bf X|Y,Z}) & = &  -\Im \left[ {\bf \{  X \cdot u(Z) \}  \{ u (Y) \cdot  u(X) \} } \right]-\Im \left[ {\bf \{ X \cdot u(Y) \}  \{ u (Z) \cdot  u(X) \} } \right] ,    \label{eq:MHD_ET:Suu} \nonumber \\ \\
  S^{bb}({\bf X|Y,Z}) & = &  - \Im \left[ {\bf \{  X \cdot u(Z) \}  \{ B (Y) \cdot  B(X) \} } \right]- \Im \left[ {\bf \{  X \cdot u(Y) \}  \{ B (Z) \cdot  B(X) \} } \right] ,  \label{eq:MHD_ET:Sbb}   \nonumber \\ \\
   S^{ub}({\bf X|Y,Z}) & = &  \Im \left[ {\bf \{  X \cdot B(Z) \}  \{ B (Y) \cdot  u(X) \} } \right] + \Im \left[ {\bf \{  X \cdot B(Y) \}  \{ B (Z) \cdot  u(X) \} } \right] ,   \label{eq:MHD_ET:Sub}  \nonumber \\ \\
    S^{bu}({\bf X|Y,Z}) & = &  \Im \left[ {\bf \{  X \cdot B(Z) \}  \{ u (Y) \cdot  B(X) \} } \right] + \Im \left[ {\bf \{  X \cdot B(Y) \}  \{ u (Z) \cdot  B(X) \} } \right] .  \label{eq:MHD_ET:Sbu}  \nonumber \\ 
 \eea
 These are the {\em combined energy transfers} to wavenumber ${\bf X}$ from the other two wavenumbers ${\bf Y}$ and ${\bf Z}$, and they are summarised in Table~\ref{tab:MHD_ET:combinedET}. For these functions, the first argument is the receiver wavenumber, while the rest two are giver wavenumbers. 
  
 \begin{table}[h]
 \caption{ Summary of various combined energy transfers in MHD turbulence.  }
 \centering
{\begin{tabular}{ c | c |  c } \hline
ET  &  Receiver mode & Giver modes     \\ \hline
$S^{uu}({\bf X|Y,Z})$ & ${\bf u(X)}$  & ${\bf u(Y)}$, ${\bf u(Z)}$      \\ \hline    
$S^{bb}({\bf X|Y,Z})$ & ${\bf B(X)}$  &  ${\bf B(Y)}$, ${\bf B(Z)}$    \\ \hline    
$S^{ub}({\bf X|Y,Z})$& ${\bf u(X)}$  &  ${\bf B(Y)}$, ${\bf B(Z)}$  \\ \hline    
$S^{bu}({\bf X|Y,Z})$ & ${\bf B(X)}$  & ${\bf u(Y)}$,  ${\bf u(Z)}$     \\ \hline  
 \end{tabular}}
\label{tab:MHD_ET:combinedET}
\end{table}

\subsection{Mode-to-mode energy transfers:}

An important question is whether we can compute the individual energy transfers to wavenumber ${\bf X}$ from wavenumbers ${\bf Y}$ and ${\bf Z}$. Dar {\em et al.}~\cite{Dar:PD2001}, Verma~\cite{Verma:PR2004}, and Verma~\cite{Verma:book:ET} derived formulas for these transfers. We will describe these formulas in the following discussion.

We denote the desired ``mode-to-mode energy transfer" using $S^{fg}({\bf X|Y|Z})$ that represents the energy transfer from mode ${\bf g(Y)}$ to mode ${\bf f(X)}$ with mode ${\bf h(Z)}$ acting as a mediator.  Thus, the receiver, giver, and mediator fields are ${\bf f, g, h}$ respectively, and the corresponding wave numbers are ${\bf X, Y, Z}$ respectively.  The receiver, giver, and mediator wavenumbers appear as arguments of $S$ in the same order.    

 \begin{figure}[htbp]
\begin{center}
\includegraphics[scale=0.4]{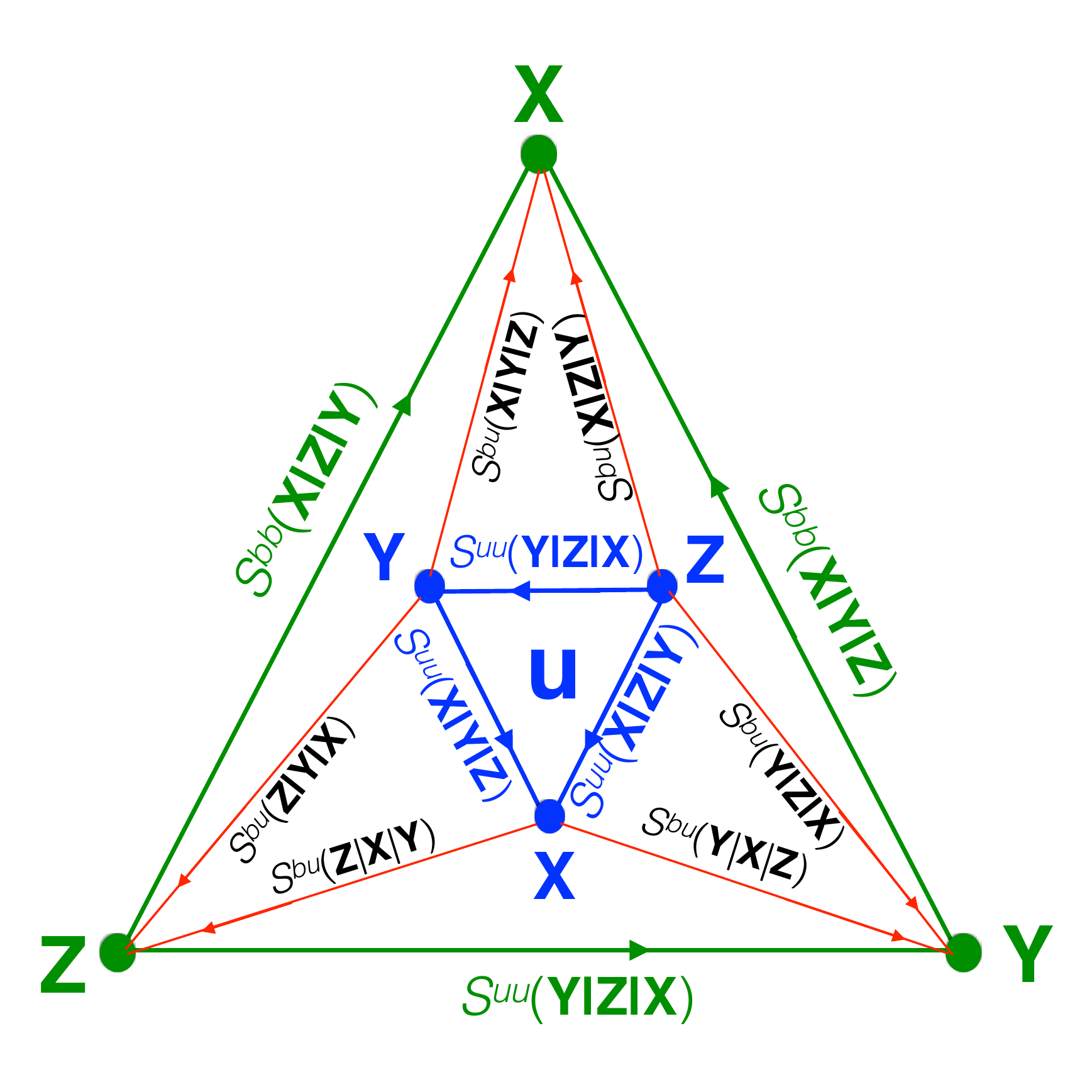}
\caption{   A wavenumbers triad $({\bf X, Y, Z})$ with  ${\bf X+Y+Z} = 0$ with blue and green symbols representing the velocity and magnetic modes respectively.  In the figure, $S^{uu}$, $S^{bb}$, $S^{bu}$ are the mode-to-mode $U2U$, $B2B$, and $U2B$  transfers. }
\label{fig:MHD_ET:MHD_S}
\end{center}
\end{figure}

In Fig.~\ref{fig:MHD_ET:MHD_S} we illustrate some of the energy transfers of MHD transfers.  From the structure of Eqs.~(\ref{eq:MHD_ET:Euk_triad}-\ref{eq:MHD_ET:Sbu}) we deduce that MHD turbulence has velocity-to-velocity ($U2U$), magnetic-to-magnetic ($B2B$), magnetic-to-velocity ($B2U$), and velocity-to-magnetic ($U2B$) energy transfers.  

The energy transfer is scalar and additive quality, similar to the financial transactions.  In addition, the energy exchanges must satisfy the following properties:
\begin{enumerate}
\item  The sum of $S^{fg}(\mathbf{X|Y|Z})$ and $S^{fg}(\mathbf{X|Z|Y})$ is the combined energy transfer $S^{fg}(\mathbf{X|Y,Z})$. That is,
\be
 S^{fg}({\bf X|Y|Z}) + S^{fg}({\bf X|Z|Y}) = S^{fg}({\bf X|Y,Z}). 
 \label{eq:MHD_ET:Sfg_sum}
 \ee
 
 \item The energy transfer from  ${\bf g}({\bf Y})$ to  ${\bf f}({\bf X})$, $S^{fg}(\mathbf{X|Y|Z})$, is  equal and opposite to the enstrophy transfer from  $ {\bf f}({\bf X})$ to ${\bf g}({\bf Y})$, $S^{gf}(\mathbf{Y|X|Z})$, i.e.,
 \be
 S^{fg}({\bf X|Y|Z}) = - S^{gf}({\bf Y|X|Z}).
  \label{eq:MHD_ET:S_equal_opposite}
 \ee
 \end{enumerate}
 In addition to the above properties, the mode-to-mode transfer functions must also satisfy Eqs.~(\ref{eq:MHD_ET:Euk_triad}-\ref{eq:MHD_ET:Sbu}). It is easy to observe that the following functions satisfy the above constraints \cite{Dar:PD2001, Verma:PR2004}: 
  \bea
 S^{uu}(\mathbf{X|Y|Z}) & = &    - \Im \left[ {\bf  \{  X \cdot u(Z) \} 
\{ u(Y) \cdot u(X) \} } \right] ,  \label{eq:MHD_ET:Suu_M2M} \\
 S^{bb}(\mathbf{X|Y|Z}) & = &    - \Im \left[ {\bf  \{  X\cdot u(Z) \} 
\{ B(Y) \cdot B(X) \} } \right] ,  \label{eq:MHD_ET:Sbb_M2M}  \\
S^{ub}(\mathbf{X|Y|Z}) & = &    - \Im \left[ {\bf  \{  X \cdot B(Z) \} 
\{ B(Y) \cdot u(X) \} } \right] , \label{eq:MHD_ET:Sub_M2M}   \\
S^{bu}(\mathbf{X|Y|Z}) & = &    - \Im \left[ {\bf  \{  X \cdot B(Z) \} 
\{ u(Y) \cdot B(X) \} } \right]  . \label{eq:MHD_ET:Sbu_M2M}
\eea
They represent the velocity-to-velocity, magnetic-to-magnetic, magnetic-to-velocity, and velocity-to-magnetic energy transfers respectively.  The formulas of Eqs.~(\ref{eq:MHD_ET:Suu_M2M}- \ref{eq:MHD_ET:Sbu_M2M}) are products of (a) a scalar product of the giver and receiver modes, and (b) a scalar product of the mediator  mode with the  receiver wavenumber.    We summarise these transfers in Table~\ref{tab:MHD_ET:M2M}.
 \begin{table}[h]
 \caption{ Summary of various mode-to-mode energy transfers in MHD turbulence.  Here the wavenumbers of the receiver, giver, and mediator modes are ${\bf X, Y, Z}$ respectively. }
{\begin{tabular}{ c | c | c | c | c } \hline
ET  &  Receiver & Giver & Mediator & Formula   \\ \hline
$S^{uu}({\bf X|Y|Z})$ & ${\bf u(X)}$  & ${\bf u(Y)}$  & ${\bf u(Z)}$  &$ -\Im \left[ {\bf  \{  X \cdot u(Z) \} \{ u(Y) \cdot u(X) \} } \right]$   \\ \hline    
$S^{bb}({\bf X|Y|Z})$ & ${\bf B(X)}$  & ${\bf B(Y)}$  & ${\bf u(Z)}$  &$ -\Im \left[ {\bf  \{ X \cdot u(Z) \} \{ B(Y) \cdot B(X) \} } \right]$   \\ \hline    
$S^{ub}({\bf X|Y|Z})$& ${\bf u(X)}$  & ${\bf B(Y)}$  & ${\bf B(Z)}$  &$ \Im \left[ {\bf  \{  X \cdot B(Z) \} \{ B(Y) \cdot u(X) \} } \right]$   \\ \hline    
$S^{bu}({\bf X|Y|Z})$ & ${\bf B(X)}$  & ${\bf u(Y)}$  & ${\bf B(Z)}$  &$ \Im \left[ {\bf  \{  X \cdot B(Z) \} \{ u(Y) \cdot B(X) \} } \right]$   \\ \hline    
\end{tabular}}
\label{tab:MHD_ET:M2M}
\end{table}

However, there is a complication.  The above solution set based on conditions of  Eqs.~(\ref{eq:MHD_ET:Euk_triad}-\ref{eq:MHD_ET:Sbu}) and Eqs.~(\ref{eq:MHD_ET:Sfg_sum},  \ref{eq:MHD_ET:S_equal_opposite}) is not unique~\cite{Dar:PD2001, Verma:PR2004}.   We invoke the following physical arguments to show that the above functions are indeed the desired mode-to-mode energy transfers.

\begin{figure}[htbp]
\begin{center}
\includegraphics[scale=0.30]{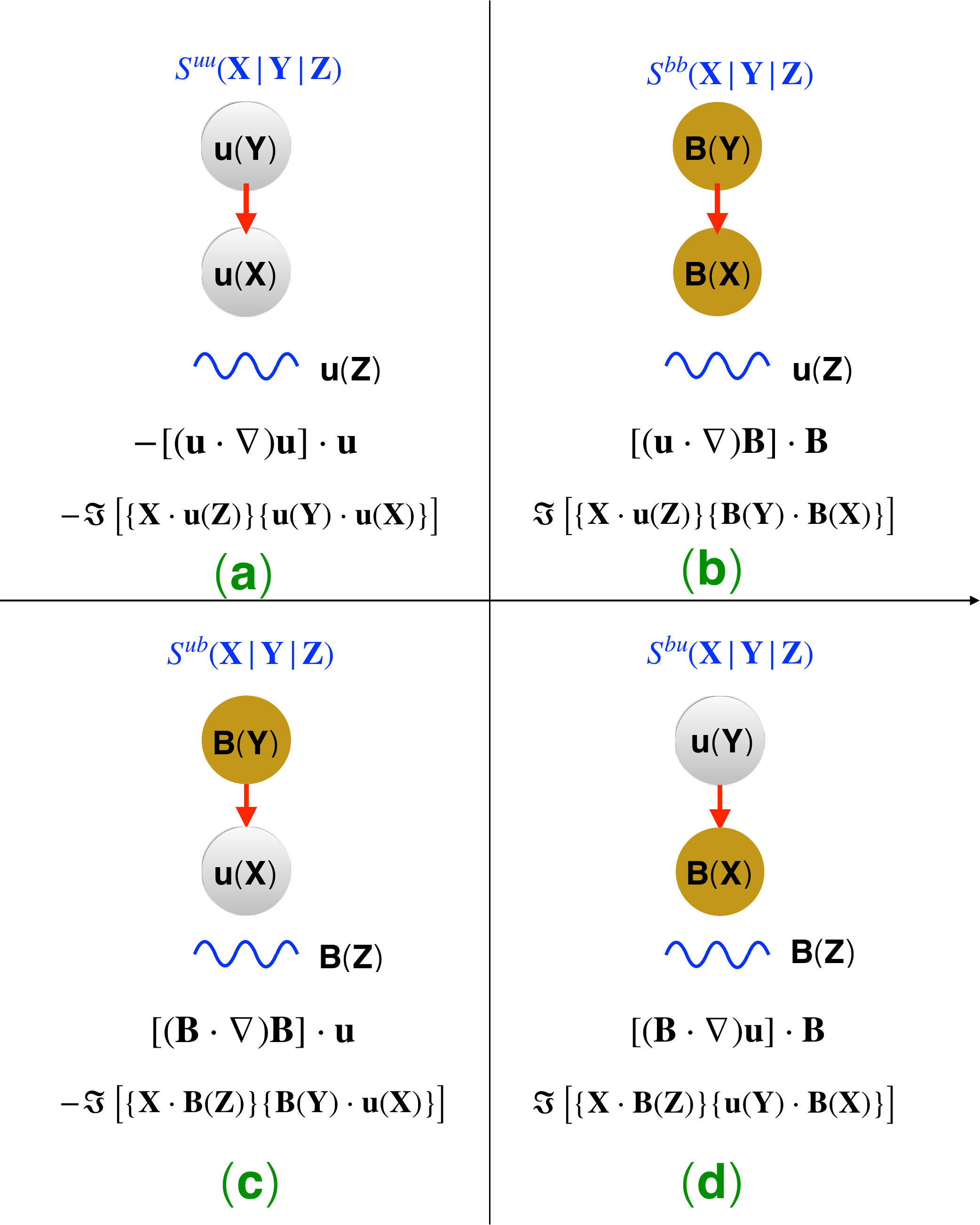}
\caption{   A schematic diagram exhibiting mode-to-mode energy transfers in MHD turbulence: (a) $U2U$, (b) $B2B$, (c) $B2U$, and (d) $U2B$.  Here the wavy lines represent the mediator modes. }
\label{fig:MHD_ET:ET_schematic_MHD}
\end{center}
\end{figure}

In Eq.~(\ref{eq:MHD_formalism:MHDB}), the nonlinear term ${\bf (u \cdot\nabla) B}$ represents the advection of the magnetic field ${\bf B}$ by the velocity field ${\bf u}$.  Therefore,  for this process, ${\bf u}$  only mediates the $B2B$ energy transfer.  In the equation for the magnetic energy  [Eq.~(\ref{eq:MHD_formalism:Eb_dynamics})], the  $B2B$ energy transfer is via the term $u_j  \partial_j  \left(\frac{1}{2} B_i B_i  \right)  $.  Here,  ${\bf u}$, which is to the left of the derivative operator,  acts as a mediator for the energy transfer between ${\bf B}$'s, which are  in the right of the derivative operator.   In Fourier space, it corresponds to $ S^{bb}(\mathbf{X|Y|Z})$ of Eq.~(\ref{eq:MHD_ET:Sbb_M2M}), which  is the energy transfer from $ {\bf B} ({\bf Y})$ to  ${\bf B}({\bf X})$ with the mediation of ${\bf u(Z)}$ (wavy line).  See Fig.~\ref{fig:MHD_ET:ET_schematic_MHD}(b) for an illustration.

Following the same arguments we can show that Eq.~(\ref{eq:MHD_ET:Suu_M2M}), which is illustrated in  Fig.~\ref{fig:MHD_ET:ET_schematic_MHD}(a), represents the $U2U$ energy transfer.  When we relate the above with the nonlinear term $u_j  \partial_j  \left(\frac{1}{2} u_i u_i  \right)  $ of Eq.~(\ref{eq:MHD_formalism:Eu_dynamics}), 
The mediator mode ${\bf u(Z)}$ corresponds to $u_j$ [as in Eq.~(\ref{eq:MHD_ET:Sbb_M2M})], while the giver and receiver modes are ${\bf u(Y)}$ and ${\bf u(X)}$ respectively.

For the other two energy transfers of Eqs.~(\ref{eq:MHD_ET:Sub_M2M}, \ref{eq:MHD_ET:Sbu_M2M}), the velocity field does not act as a advector.  Rather, the magnetic field takes that role, as illustrated in Fig.~\ref{fig:MHD_ET:ET_schematic_MHD}(c,d).  For both these transfers, ${\bf B(Z)}$ acts as a mediator.  In $S^{bu}({\bf X|Y|Z})$, ${\bf B(X)}$ receives energy from ${\bf u(Y)}$, but in $S^{ub}({\bf X|Y|Z})$, ${\bf u(X)}$ receives energy from ${\bf B(Y)}$.

We remark that that Verma~\cite{Verma:book:ET} also provided symmetry-based arguments to show that  Eqs.~(\ref{eq:MHD_ET:Suu_M2M}- \ref{eq:MHD_ET:Sbu_M2M}) are indeed the respective mode-to-mode energy transfers of MHD turbulence. These derivations however are beyond the scope of this paper.

\section{Energy fluxes in MHD turbulence:}
 \label{sec:energy_fluxes}
 
\begin{figure}[htbp]
\begin{center}
\includegraphics[scale = 0.6]{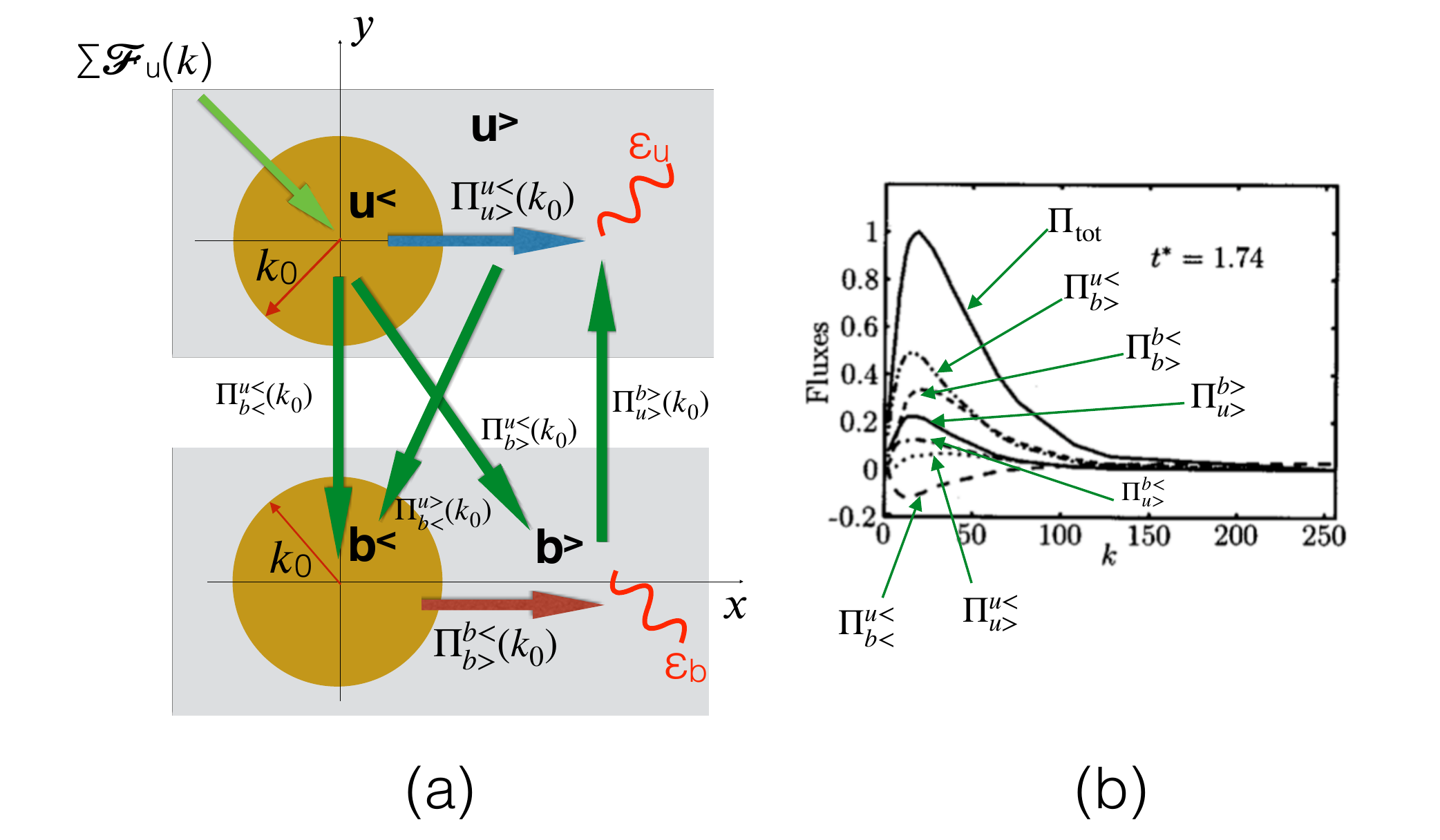}
\end{center}
\caption{(a) Various energy fluxes in MHD turbulence. (b) Energy fluxes computed by   Debliquy {\em et al.}~\cite{Debliquy:PP2005}. }
\label{fig:MHD_ET:Pi_MHD}
\end{figure}

 The mode-to-mode energy transfers provide perfect recipe for computing the energy fluxes for MHD turbulence.  We consider a wavenumber sphere of radius $k_0$. Various  energy fluxes for this sphere  are:
\begin{enumerate}
\item $\Pi^{u<}_{u>}(k_{0})$:  Energy transfers from all the velocity modes  inside the sphere to all the velocity modes outside the sphere:
\be
\Pi^{u<}_{u>}(k_{0})= \sum_{|\mathbf{Y}|\le k_{0}} \sum_{|\mathbf{X}|>k_{0}} S^{uu}({\bf X|Y|Z}).
\label{eq:MHD_ET:MHD_flux_u<_u>}
\ee

\item $\Pi^{u<}_{b>}(k_{0})$:  Energy transfers from all the velocity modes  inside the sphere to all the magnetic modes outside the sphere:
\be
\Pi^{u<}_{b>}(k_{0})= \sum_{|\mathbf{Y}|\le k_{0}} \sum_{|\mathbf{X}|>k_{0}} S^{bu}({\bf X|Y|Z}).
\label{eq:MHD_ET:MHD_flux_u<_b>}
\ee

\item $\Pi^{b<}_{b>}(k_{0})$:  Energy transfers from all the magnetic modes  inside the sphere to all the magnetic modes outside the sphere:
\be
\Pi^{b<}_{b>}(k_{0})= \sum_{|\mathbf{Y}|\le k_{0}} \sum_{|\mathbf{X}|>k_{0}} S^{bb}({\bf X|Y|Z}).
\label{eq:MHD_ET:MHD_flux_b<_b>}
\ee

\item $\Pi^{u>}_{b<}(k_{0})$:  Energy transfers from all the velocity modes  outside the sphere to all the magnetic modes inside the sphere:
\be
\Pi^{u>}_{b<}(k_{0})= \sum_{|\mathbf{Y}| > k_{0}} \sum_{|\mathbf{X}|\le k_{0}} S^{ub}({\bf X|Y|Z}).
\label{eq:MHD_ET:MHD_flux_b<_u>}
\ee

\item $\Pi^{u<}_{b<}(k_{0})$:  Energy transfers from all the velocity modes  inside the sphere to all the magnetic modes inside the sphere:
\be
\Pi^{u<}_{b<}(k_{0})= \sum_{|\mathbf{Y}|\le k_{0}} \sum_{|\mathbf{X}|\le k_{0}} S^{ub}({\bf X|Y|Z}).
\label{eq:MHD_ET:MHD_flux_u<_b<}
\ee

\item $\Pi^{u>}_{b>}(k_{0})$:  Energy transfers from all the velocity modes  outside the sphere to all the magnetic modes outside the sphere:
\be
\Pi^{u>}_{b>}(k_{0})= \sum_{|\mathbf{Y}|> k_{0}} \sum_{|\mathbf{X}| > k_{0}} S^{bu}({\bf X|Y|Z}).
\label{eq:MHD_ET:MHD_flux_u>_b>}
\ee
\end{enumerate}
In the above formulas, the superscripts and subscripts represents the giver  and  receiver  modes respectively, while $<$ and $ >$ represent the modes residing inside and outside the sphere respectively. The above fluxes are illustrated in Fig.~\ref{fig:MHD_ET:Pi_MHD}(a). 

In Fig.~\ref{fig:MHD_ET:Pi_MHD}(b) we illustrate the energy fluxes computed by Debliquy {\em et al.}~\cite{Debliquy:PP2005} using numerical data of  decaying  turbulence simulation. Some important observations from the plot are: (a) The $B2B$ energy transfer is positive; (b) There is an energy transfer from the large-scale magnetic field ($b<$ ) to large-scale velocity field $(u<)$; (3) The $U2U$ transfer is strongly suppressed. Note however that the energy transfers for forced MHD and decaying MHD are quite different. For example, forced MHD always exhibit positive $\Pi^{u<}_{b<}(k)$.

An important regime of MHD turbulence is steady state, which is obtained when the energy supply by ${\bf F}_\mathrm{ext}$ matches with the total dissipation rate $\epsilon_\mathrm{tot}$.  We denote the viscous dissipation rate of the velocity field by $\epsilon_u$, and the Joule dissipation rate of the magnetic field by $\epsilon_b$ (see Fig.~\ref{fig:MHD_ET:Pi_MHD}(a)).  Note that  $\epsilon_\mathrm{tot} =  \epsilon_u+\epsilon_b$.

Here we state without proof some of the identities for a wavenumber sphere in the inertial range:
  \bea
\Pi^{u<}_{u>}(k) + \Pi^{u<}_{b>}(k) + \Pi^{b<}_{b>}(k) + \Pi^{b<}_{u>}(k) & = &  \Pi_\mathrm{tot}(k) = \epsilon_\mathrm{tot}, \\
\Pi^\mathrm{all}_{u>}(k)=   \Pi^{u<}_{u>}(k) + \Pi^{b<}_{u>}(k) + \Pi^{b>}_{u>}(k) & = & \epsilon_u, \label{eq:MHD_ET:Piu_conservation} \\
\Pi^\mathrm{all}_{b>}(k)=     \Pi^{b<}_{b>}(k) + \Pi^{u<}_{b>}(k) + \Pi^{u>}_{b>}(k) & = &   \epsilon_b,
    \label{eq:MHD_ET:Pib_conservation} \\
     \Pi^{u<}_{b<}(k) + \Pi^{u<}_{b>}(k) + \Pi^{u>}_{b<}(k) + \Pi^{u>}_{b>}(k) &= & \epsilon_b \\
  \Pi^{u<}_{b<}(k) + \Pi^{u>}_{b<}(k) & = & \Pi^{b<}_{b>}(k), \\
    \Pi^{u<}_{u>}(k)  + \Pi^{u<}_{b<}(k)  + \Pi^{u<}_{b>}(k)  & = & \sum_{\bf k} \mathcal{F}_\mathrm{ext}({\bf k}) = \epsilon_\mathrm{tot}.
   \eea
Note that the above identities are not all independent.

In the next section we describe how the energy fluxes can provide valuable inputs to the dynamo process, which is the magnetic energy growth in astrophysical objects.

\section{Application to large-scale dynamo:}
\label{sec:dynamo}

The dynamo process remains an unsolved problem~\cite{Moffatt:book}. Researchers have attempted to understand dynamo process by various schemes: $\alpha$-dynamo,  experimental dynamos, dynamo transition~\cite{Yadav:EPL2010}, numerical dynamos, etc.  Also, there are   many parameters---Prandtl number, Reynolds number, geometry, forcing, etc.---that affect the dynamo process.  

In this section, we present energy fluxes for large-scale dynamo that is forced at intermediate scale.  An important question is whether the large-scale magnetic field will grow.    Kumar and Verma~\cite{Kumar:PP2017} studied one such dynamo in which the forcing was employed at scale  1/10 the box size, i.e.,  $k \in (10,11)$.  They performed numerical simulation of  this dynamo, and then analysed various energy fluxes at different stages of the magnetic field growth.  The flow was forced with a net kinetic energy injection rate of unity but with negligible kinetic and magnetic helicity. Thus, this is a nonhelical dynamo.   Also, the magnetic Prandtl number Pm was chosen to be unity.

   \begin{figure}[htbp]
\begin{center}
\includegraphics[scale = 0.6]{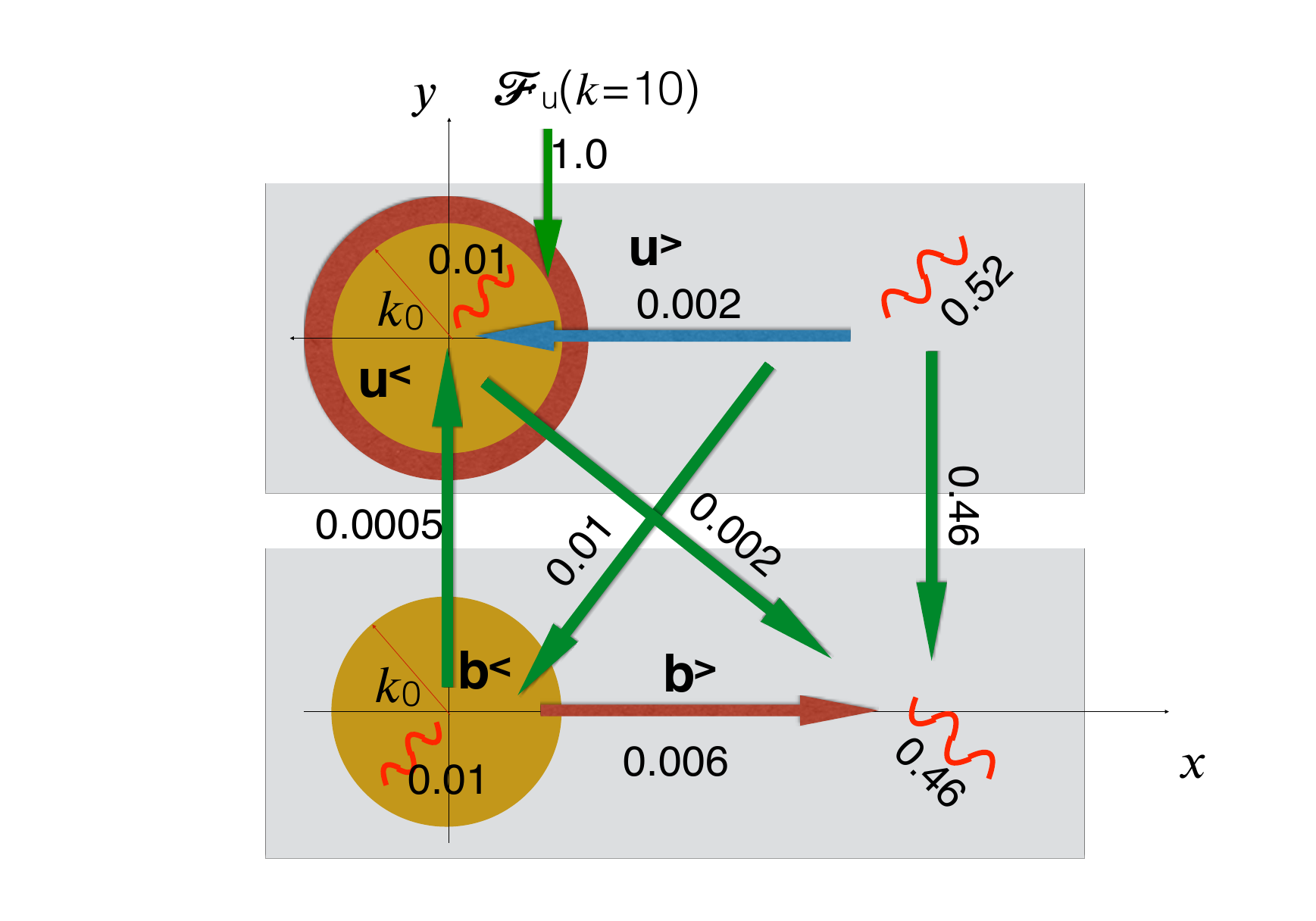}
\end{center}
\caption{Various energy fluxes for the nonhelical dynamo of Kumar and Verma~\cite{Kumar:PP2017}.  These fluxes are of $t=127$ eddy turnover time.  }
\label{fig:MHD_ET:dyn_flux}
\end{figure}

In Fig.~\ref{fig:MHD_ET:dyn_flux} we exhibit various energy fluxes and dissipation rates  towards the later stages ($t=127$ eddy turnover time) of the dynamo process when kinetic and magnetic energies have reached a steady state.    The central brown circle in Fig.~\ref{fig:MHD_ET:dyn_flux} represent sphere of radius $k_0 =8$.  Note that $k_0 < k_f$, thus it represents the large scales.  When we focus on this figure, we observe that  $\Pi^{u>}_{b<}(k_0) \approx 0.01$ is the most dominant  energy input to this sphere.  Also, the $b<$ sphere loses relatively small amount of energy to $u<$ and $b>$.  The Joule dissipation in the sphere,  represented by red wavy line, appears to balance the energy input to the sphere.  Thus, the large-scale magnetic field is sustained by the  $\Pi^{u>}_{b<}$ energy flux. This energy flux is only 1\% of the total energy injection rate, but this is sufficient because the Joule dissipation is quite weak due to the $k^2$ factor. 

\begin{figure}[htbp]
\begin{center}
\includegraphics[scale = 0.3]{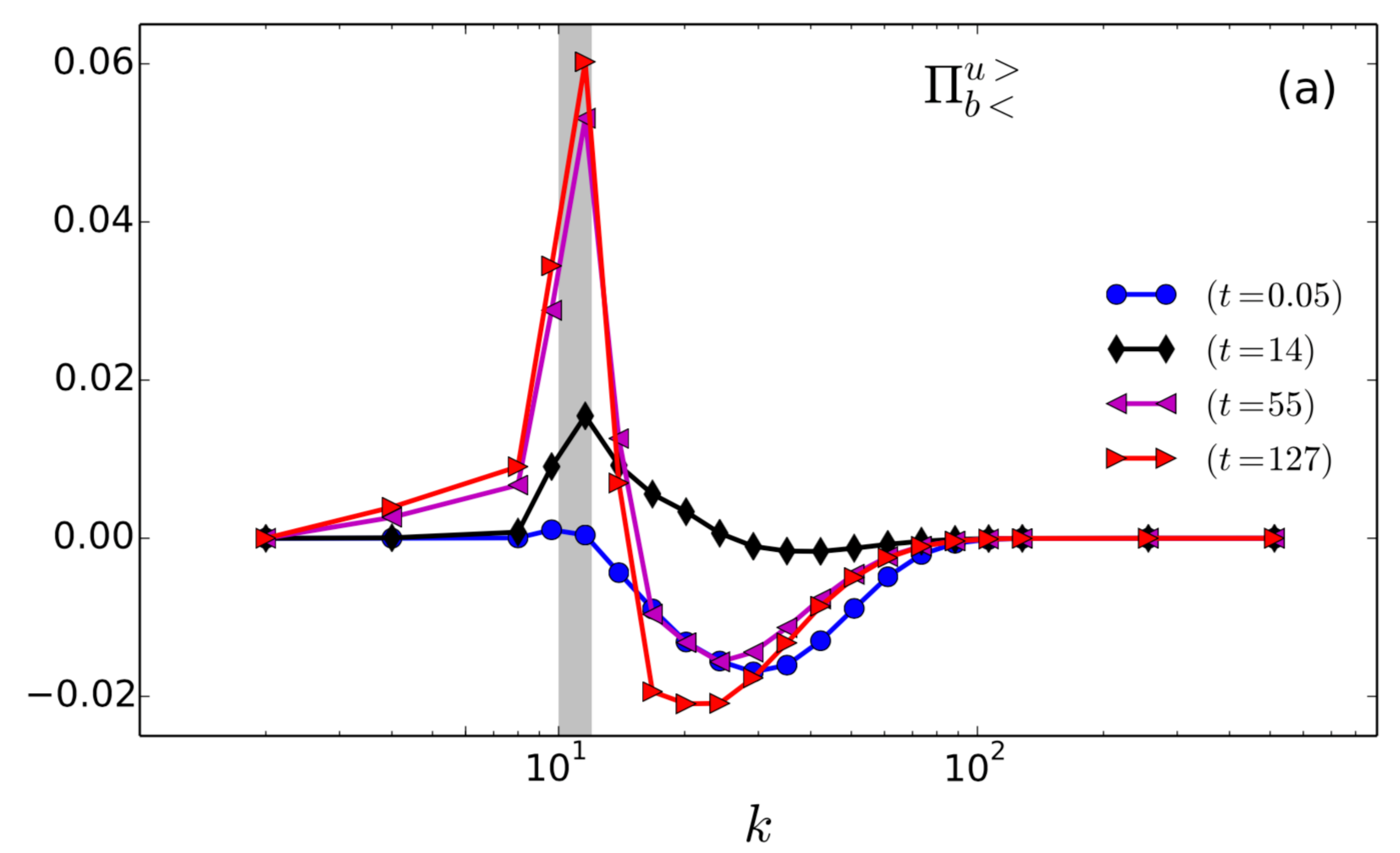}
\end{center}
\caption{The energy fluxes $\Pi^{u>}_{b<}$ in MHD turbulence at various stages of the dynamo.   All the energy fluxes of Fig.~\ref{fig:MHD_ET:dyn_flux} are for $t=127$  eddy turnover time. }
\label{fig:MHD_ET:u_gtr_b_less}
\end{figure}

This dynamo has  other  interesting features, e.g, $\epsilon_u \approx 0.53$ (sum of 0.52 and 0.01) and $\epsilon_b =0.47$ at $t=127$ time unit. Figure~\ref{fig:MHD_ET:u_gtr_b_less} illustrates plots of energy flux $\Pi^{u>}_{b<}$  at other instances.  These plots show that large-scale magnetic field always get small amount of energy from the forcing band.  

Kumar {\em et al.}~\cite{Kumar:EPL2014,Kumar:JoT2015} analysed energy transfers for small and large magnetic Prandtl numbers and reported the dominant sources of inputs to the magnetic energy. These computations show that the energy fluxes provide valuable insights into the dynamo mechanism.

%

\section{Conclusions:}
\label{sec:conclusions}

In this paper we report various energy transfers in MHD turbulence.  In particular, we detailed the mode-to-mode velocity-to-velocity, velocity-to-magnetic, magnetic-to-magnetic, and magnetic-to-velocity energy transfers.  Then, we described the six energy fluxes of MHD turbulence that provide valuable information on the energy transfers among large-scale and small-scale fields.

The above energy fluxes and other related quantities are very useful for understanding turbulence dynamics. In this paper we illustrate how the energy fluxes help us understand the growth of the large-scale magnetic field when the external forcing is employed at the intermediate scale.  

We also remark that the above formalism of energy transfers is quite general, and it has been extended to other systems such as passive scalar, thermal convection~\cite{Verma:book:BDF}, shell model~\cite{Verma:JoT2016,Plunian:PR2012}, etc.  Dar \etal~\cite{Dar:PD2001} and Verma~\cite{Verma:PR2004} also showed how the mode-to-mode energy transfer formalism removes the ambiguity in shell-to-shell energy transfer computations encountered in the calculations that employ the formalism of combined energy transfer.

\section*{Acknowledgements}
Many collaborators contributed to  the investigation of energy transfers in MHD turbulence and dynamo.  We thank them, in particular Gaurav Dar, V. Eswaran, and Rohit Kumar,  for this collaboration.  This work was supported by the research grants PLANEX/PHY/2015239 from Indian Space Research Organisation, India, and  by the  Department of Science and Technology, India (INT/RUS/RSF/P-03) and Russian Science Foundation Russia (RSF-16-41-02012) for the Indo-Russian project.  The dynamo simulations were performed on Shaheen II of the Supercomputing Laboratory at King Abdullah University of Science and Technology (KAUST) under the project K1052, and on Chaos supercomputer of Simulation and Modeling Laboratory (SML), IIT Kanpur.




%
%





\lastpageno
\end{document}